\documentclass[sn-nature]{sn-jnl}
\usepackage[T1]{fontenc}
\usepackage{graphicx}%
\usepackage{multirow}%
\usepackage{amsmath,amssymb,amsfonts}%
\usepackage{amsthm}%
\usepackage{mathrsfs}%
\usepackage[title]{appendix}%
\usepackage{xcolor}%
\usepackage{textcomp}%
\usepackage{manyfoot}%
\usepackage{booktabs}%
\usepackage{algorithm}%
\usepackage{algorithmicx}%
\usepackage{algpseudocode}%
\usepackage{listings}%

\theoremstyle{thmstyleone}%
\theoremstyle{thmstyletwo}%

\theoremstyle{thmstylethree}%

\begin{document}

\title[Article Title]{Gate Stack Engineering for High-Mobility and Low-Noise SiMOS Quantum Devices}

%%=============================================================%%
%% Prefix	-> \pfx{Dr}
%% GivenName	-> \fnm{Joergen W.}
%% Particle	-> \spfx{van der} -> surname prefix
%% FamilyName	-> \sur{Ploeg}
%% Suffix	-> \sfx{IV}
%% NatureName	-> \tanm{Poet Laureate} -> Title after name
%% Degrees	-> \dgr{MSc, PhD}
%% \author*[1,2]{\pfx{Dr} \fnm{Joergen W.} \spfx{van der} \sur{Ploeg} \sfx{IV} \tanm{Poet Laureate} 
%%                 \dgr{MSc, PhD}}\email{iauthor@gmail.com}
%%=============================================================%%

\author*[1]{\fnm{Md. Mamunur} \sur{Rahman}}
\email{md\_mamunur.rahman@unsw.edu.au}
%\equalcont{These authors contributed equally to this work.}

%\equalcont{These authors contributed equally to this work.}

\author[1,2]{\fnm{Ensar} \sur{Vahapoglu}}
\author[1,2]{\fnm{Kok Wai} \sur{Chan}}
\author[1,2]{\fnm{Tuomo} \sur{Tanttu}}
\author[1,2]{\fnm{Ajit} \sur{Dash}}
\author[4]{\fnm{Jonathan} \sur{Yue Huang}}
\author[1,2,5]{\fnm{Steve} \sur{Yianni}}
\author[1]{\fnm{Venkatesh} \sur{Chenniappan}}
\author[1,2]{\fnm{Jesús D.} \sur{Cifuentes}}
\author[1,2]{\fnm{Fay} \sur{Hudson}}
\author[1,2]{\fnm{Christopher} \sur{C. Escott}}
\author[2,3]{\fnm{Yik Kheng} \sur{Lee}}
\author[1,2]{\fnm{Nard Dumoulin} \sur{Stuyck}}
\author[1,2]{\fnm{Arne} \sur{Laucht}}
\author[1]{\fnm{Andrea} \sur{Morello}}
\author[1,2]{\fnm{Andre} \sur{Saraiva}}
%\author[4]{\fnm{Alexander} \sur{Hamilton}}
\author[2,3]{\fnm{Jared H} \sur{Cole}}
\author[1,2]{\fnm{Andrew S.} \sur{Dzurak}}
%\email{a.dzurak@unsw.edu.au}
\author*[1,2]{\fnm{Wee Han} \sur{Lim}}\email{wee.lim@unsw.edu.au}

\affil[1]{\orgdiv{School of Electrical Engineering and Telecommunications}, \orgname{The University of New South Wales}, \city{Sydney}, \postcode{2052}, \state{New South Wales}, \country{Australia}}

\affil[2]{\orgname{Diraq}, \city{Sydney}, \postcode{2052}, \state{New South Wales}, \country{Australia}}

\affil[3]{\orgdiv{School of Science}, \orgname{RMIT University}, \city{Melbourne}, \postcode{3000}, \state{Victoria}, \country{Australia}}

\affil[4]{\orgdiv{School of Physics}, \orgname{The University of New South Wales}, \city{Sydney}, \postcode{2052}, \state{New South Wales}, \country{Australia}}
\affil[5]{\orgname{Current Address: CEA- Interdisciplinary Research Institute of Grenoble}, \city{Grenoble}, \postcode{38054}, \country{France}}

%%==================================%%
%% sample for unstructured abstract %%
%%==================================%%

\abstract{

We systematically investigate the interplay between materials engineering, quantum transport, and low-frequency charge noise in silicon metal–oxide–semiconductor (SiMOS) quantum devices. By combining Hall-bar transport measurements with charge-noise spectroscopy of gate-defined quantum dots, we identify correlations between gate-stack design, carrier mobility, and electrostatic noise, providing an experimental case study of material and process dependencies relevant to low-noise, high-mobility operation. Hall-bar studies reveal that increasing the atomic-layer-deposition temperature of Al$_2$O$_3$ markedly enhances mobility, whereas the choice of oxidant has little impact. Devices incorporating HfO$_2$ exhibit improved carrier mobility, an interesting observation that can plausibly be attributed to defect passivation associated with aluminum diffusion from the gate metal into the HfO$_2$ layer. Charge-noise measurements show a strong correlation between higher mobility and reduced noise, with TiPd-gated devices displaying both degraded transport and elevated charge noise. In contrast, the poly-Si-gated CMOS-foundry device achieves the lowest noise levels. Finally, dual-feedback dot–sensor stability mapping demonstrates enhanced charge stability in devices with the gate stacks studied here, underscoring their promise for scalable, high-fidelity silicon spin-qubit platforms.

}


\maketitle
\section{Introduction}\label{sec 1}

Electron spin qubits confined in electrostatically-defined silicon quantum dots constitute a leading platform for semiconductor-based quantum computing~\cite{Quantum_computation_with_quantum_dots, Silicon_quantum_electronics}. This approach combines nanoscale device footprints~\cite{Vandersypen2019QuantumCW}, long spin coherence times~\cite{A_two_qubit_logic_gate_in_silicon,Stano2022}, and direct compatibility with advanced semiconductor manufacturing technologies~\cite{Maurand2016,george202412,stuyck2024cmos,Zwerver2022}. Over the past decade, rapid experimental progress has established single-qubit gate fidelities above 99.9\%~\cite{Silicon_qubit_fidelities_approaching_incoherent_noise_limits_via_pulse_engineering, A_quantum_dot_spin_qubit_with_coherence_limited_by_charge_noise_and_fidelity,wu2025}, two-qubit gate fidelities exceeding 99\%~\cite{Steinacker2024,Two-qubit_silicon_quantum_processor_with_operation_fidelity_exceeding,Noiri2022}, and robust qubit operation at temperatures above 1~K~\cite{Operation_of_a_silicon_quantum_processor_unit_cell_above_one_kelvin,Petit2020_UniversalQuantumLogic,High-fidelity_spin_qubit_operation_and_algorithmic_initialization_above_1K}. These milestones underscore the strong potential of silicon quantum-dot platforms for scalable quantum information processing. Nevertheless, the realization of fault-tolerant quantum computing remains challenging because quantum error correction imposes a substantial resource overhead, typically requiring hundreds to thousands of physical qubits to encode a single logical qubit~\cite{Fowler2012, Terhal2015}. In semiconductor spin-qubit platforms, the error-correction overhead is largely driven by gate errors arising from charge noise, namely, low-frequency electrical fluctuations originating from the surrounding material environment~\cite{Silicon_qubit_fidelities_approaching_incoherent_noise_limits_via_pulse_engineering,A_quantum_dot_spin_qubit_with_coherence_limited_by_charge_noise_and_fidelity,Connors2019,Struck2020,Connors2022}.

Charge noise in semiconductor quantum dots is commonly attributed to microscopic defects in the surrounding material environment, including traps at the Si/SiO$_2$ interface and within the gate oxide~\cite{PaqueletWuetz2023,Spruijtenburg2016,Culcer2009,Kepa2023}. These defects act as bistable two-level fluctuators (TLFs) that stochastically switch between distinct charge or configurational states, generating random telegraph signals. Each fluctuator produces a Lorentzian noise spectrum, and an ensemble with a broad distribution of switching rates naturally gives rise to $1/f$ noise. This behavior is described by the McWhorter model~\cite{McWhorter1957} and is consistent with the Dutta--Horn framework for thermally activated fluctuators~\cite{DuttaHorn1981}. Although originally developed for classical devices, this framework extends to quantum dots, where environmental charge fluctuations modulate the local electrostatic confinement potential~\cite{Connors2022}. Despite strong experimental support for this TLF-based picture, the precise atomic-scale nature of the underlying defects remains unresolved, underscoring the need for improved materials and interface engineering to suppress charge noise and enable scalable, high-fidelity qubit operation.

%Low temperature electron mobility of Hall bar devices is one of the typically used metric to understand the impact of material engineering in silicon metal-oxide-semiconductor (SiMOS) based system~\cite{Quantum_Transport_Properties_of_Industrial, High_mobility_SiMOSFETs_fabricated_in_a_full_300_mm_CMOS_process,wendoloski2025holessiliconheavierexpected}. Besides, percolation threshold density, and quantum lifetime provides valuable insights about the interface disorder. In this work, we initiated with microscopic Hall bar devices with different high-k based gate oxide and gate metal selection. Based on the assessment results of these microscopic device, we fabricated nanoscale quantum dot devices with studied gate oxide and gate metal in Hall bars to observe the material effect on charge noise data. Finally, we verified our observation of charge noise data with the charge sensing measurements in double quantum dot devices. These observations provides valuable insights on the material dependency of charge noise behaviour. 

Low-temperature electron mobility in Hall-bar devices has long served as a fundamental benchmark for evaluating the influence of materials engineering and interface quality in silicon metal--oxide--semiconductor (SiMOS) systems~\cite{Quantum_Transport_Properties_of_Industrial, High_mobility_SiMOSFETs_fabricated_in_a_full_300_mm_CMOS_process,wendoloski2025holessiliconheavierexpected}. In addition, the percolation threshold density and the quantum lifetime provide complementary insights into disorder and scattering at the semiconductor--oxide interface. In this work, we first systematically characterized microscopic Hall-bar devices incorporating different high-$\kappa$ gate oxides and gate-metal stacks to establish material-dependent low-temperature transport benchmarks. Guided by these results, nanoscale quantum dot devices employing the same gate-oxide and gate-metal combinations were subsequently fabricated to directly examine the influence of materials engineering on charge-noise behavior. Finally, we corroborated the observed charge-noise trends through charge-sensing measurements performed on double quantum dot devices. Together, these measurements established a consistent connection between material choice, low-temperature transport properties, and charge-noise performance, providing quantitative insight into the material dependence of charge noise in SiMOS-based quantum devices.

\section{Methods}\label{sec11}

\begin{figure}[!htbp]
\centering
\includegraphics[width=1\textwidth]{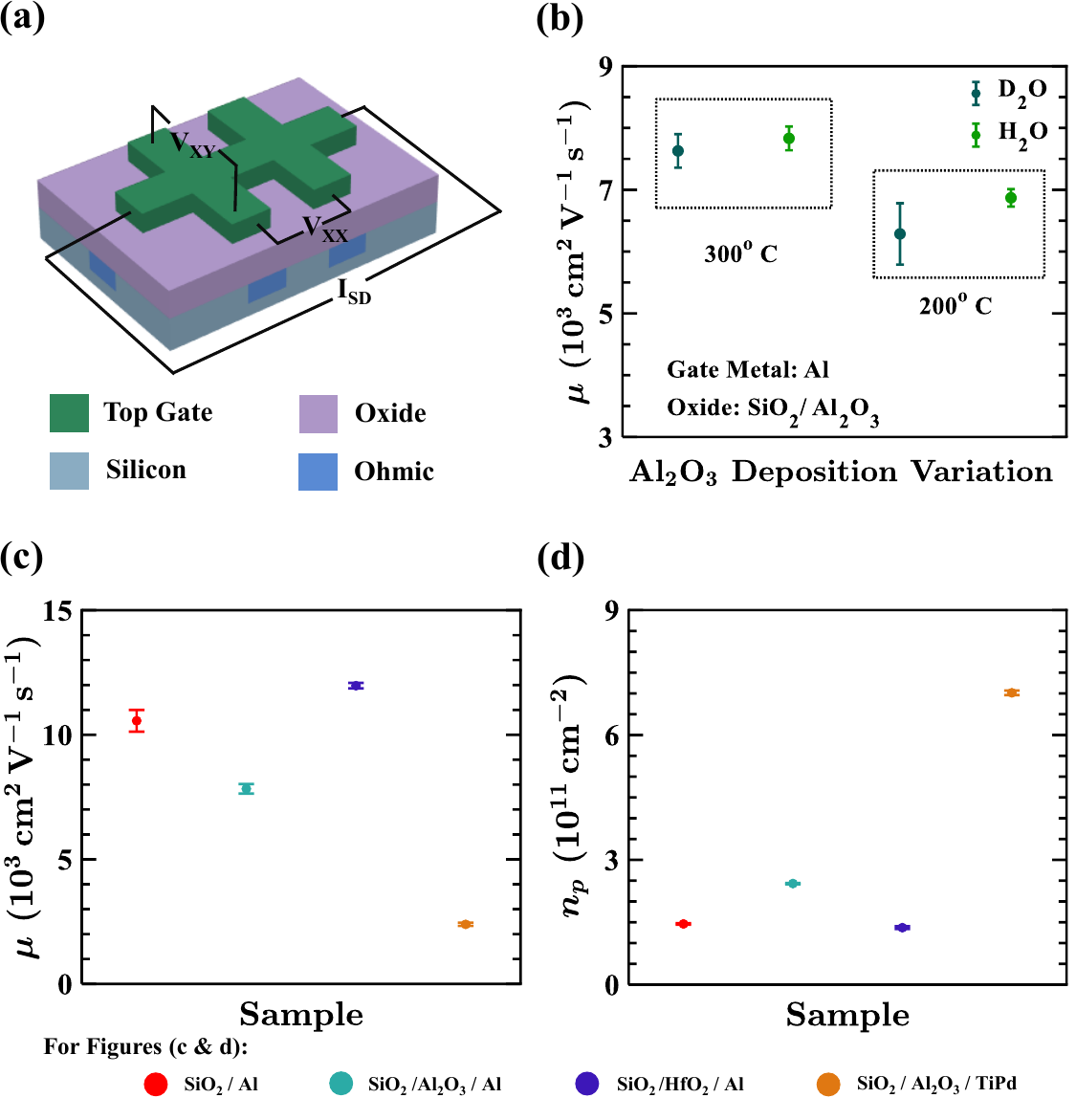}

\caption[\textbf{Hall-bar device structure and peak-mobility comparison as a function of gate-stack engineering at 1.4~K.}]{
\textbf{Hall-bar device structure and peak-mobility comparison as a function of gate-stack engineering at 1.4~K.}
\textbf{(a)} Schematic illustration of the fabricated Hall-bar device (not drawn to scale), highlighting the top gate, gate-oxide stack, and the n$^{+}$ doped ohmic contact regions.
\textbf{(b)} Peak mobility, $\mu_{\mathrm{peak}}$, for SiO$_2$/Al$_2$O$_3$-based gate stacks grown under various Al$_2$O$_3$ deposition conditions. The oxidants (H$_2$O and D$_2$O) and deposition temperatures (200 and 300$^{\circ}$C) were systematically varied to assess their impact on oxide quality and electron mobility.
\textbf{(c)} Comparison of peak mobilities from four device configurations incorporating different gate-oxide stacks and gate-metal materials. For SiO$_2$/Al$_2$O$_3$ devices, the highest-mobility condition identified in panel (b) is used for benchmarking.
\textbf{(d)} The percolation density, $n_{\mathrm{p}}$, extracted for each device type (corresponding to the mobility data in panel (c)). Devices with lower mobility consistently exhibit higher percolation densities, indicative of increased disorder in the conduction channel. At least two devices were measured per configuration for the data shown in panels (c) and (d). The error bars represent the standard deviation of the measured values across devices within each configuration. 
}
\label{Figure 1_Chargenoise}
\end{figure}

The MOS-based six-terminal Hall-bar devices, schematically shown in Fig.~\ref{Figure 1_Chargenoise}(a), were fabricated in the University of New South Wales (UNSW) cleanroom (UC) on high-resistivity Si wafers  ($>10~\mathrm{k}\Omega\,$cm at 300~K) containing photolithographically defined  $n$-type phosphorus-doped regions. An 8~nm SiO$_2$ gate oxide was thermally grown at 825$^{\circ}$C and subsequently subjected to rapid thermal annealing (RTA) to activate the dopants and improve the Si/SiO$_2$ interface. Ohmic contacts were formed by depositing 200~nm of aluminum (Al) via electron-beam (e-beam) evaporation onto the phosphorus-doped silicon regions. For bilayer oxide stacks, an additional $\sim$6~nm high-$\kappa$ dielectric (Al$_2$O$_3$  or HfO$_2$) was deposited using a thermal atomic layer deposition (ALD) system. Al$_2$O$_3$ films were grown using trimethylaluminum (TMA) with either H$_2$O or D$_2$O, as oxidants at either 200$^{\circ}$C or 300$^{\circ}$C, while HfO$_2$ film was deposited at 250$^{\circ}$C using tetrakis(ethylmethylamino)hafnium (TEMAH) and H$_2$O. The Hall-bar top gate, defining a 10~$\mu$m-wide and 68~$\mu$m-long channel, was patterned by photolithography. Most devices used 25~nm of thermally evaporated Al as the gate metal, whereas one variant employed a 3/22~nm TiPd stack deposited by e-beam evaporation, with the 3~nm Ti layer serving as an adhesion layer. All devices underwent a forming-gas anneal (FGA) at 400$^{\circ}$C for 15~minutes following metallization and lift-off.

%The University cleanroom(UC) double quantum dot structures, invested in this study, were also fabricated with similar high-resistivity Si wafers, as mentioned above, with multilevel gate-stack silicon metal–oxide–semiconductor technology{\cite{Gate_Defined_Quantum_Dots_in_Intrinsic_Silicon}}. Following the thermally grown 8~nm SiO$_2$ oxide and ohmic contacts formation, the first layer  Al gates was fabricated with electron-beam lithography (EBL), thermal evaporation and liftoff. The gates were then oxidized to create a 4 ~nm aluminum oxide layer for electrical isolation. The second and third layer gates were fabricated similarly by strictly maintaining the alignment with the first layer gates during EBL process. A single-electron transistor (SET) was also formed as a charge sensor to monitor the charge state of the quantum-dot system. Finally, the devices also got similar heat treatment as hallbar devices. A detail of the fabrication process is described in {\cite{Electrostatically_defined_few-electron_double_quantum_dot_in_silicon}}. 

The UC double quantum dot devices shown in Fig.~\ref{Figure 2_Chargenoise}(a) were fabricated on high-resistivity silicon wafers, similar to those described above, using a multilayer gate-stack SiMOS process~\cite{Gate_Defined_Quantum_Dots_in_Intrinsic_Silicon}. After growing an 8~nm thermal SiO$_2$ layer and forming ohmic contacts,  $\sim$6~nm high-$\kappa$ dielectrics were deposited as required. The first Al gate layer was patterned by electron-beam lithography (EBL), deposited via thermal evaporation, and defined by liftoff. These gates were then oxidized to form a $\sim$4~nm AlO$_x$ layer providing electrical insulation for subsequent gate layers. The second and third gate layers were fabricated using the same EBL, evaporation, and liftoff procedures, with precise alignment to the underlying gate structures. A single-electron transistor (SET) was incorporated as a charge sensor to monitor the charge occupancy of the double quantum dot system. For devices employing TiPd gate metals, an ALD-grown $\sim$4~nm Al$_2$O$_3$ layer was used as the inter-gate dielectric in place of thermally grown AlO$_x$. Finally, all devices received the same post-fabrication heat treatment as the Hall-bar structures. A detailed description of the fabrication process is provided in  Ref.~\cite{Electrostatically_defined_few-electron_double_quantum_dot_in_silicon}.

%The Diraq's double quantum dot devices were fabricated at imec using the research technology foundry (RTF) process on an isotopically enriched layer of silicon. A 12 nm thermally grown SiO2 layer acted as the gate oxide. A tri-layer overlapping polysilicon gate stack was then deposited using the e-beam lithography and dry etching process, where a 7-8 nm ALD SiO2 was used as inter-gate dielectric. The polysilicon gates were chosen over the metallic gates to minimise the cryogenic interface strain {\cite{Stuyck_2021, Formation_of_strain_induced_quantum_dots_in_gated_semiconductor_nanostructures}}.  A detail description odf the fabrication process is available at {\cite{Low_charge_noise_quantum_dots_with_industrial_CMOS_manufacturing, Steinacker2024}}. 

The research technology foundry (RTF) double quantum dot devices shown in Fig.~\ref{Figure 2_Chargenoise}(b) were fabricated at imec on an isotopically enriched silicon layer using the RTF process. The device structure consists of a 12~nm thermally grown SiO$_2$ gate oxide, above which a tri-layer overlapping polysilicon (poly-Si) gate stack is patterned using EBL and dry etching. Inter-gate insulation was provided by a 7--8~nm ALD-grown SiO$_2$ layer. Poly-Si gates were chosen instead of metallic gates to reduce cryogenic strain at the gate--oxide interface~\cite{Stuyck2021, Formation_of_strain_induced_quantum_dots_in_gated_semiconductor_nanostructures}. Further details of the fabrication process are available in Ref.~\cite{Low_charge_noise_quantum_dots_with_industrial_CMOS_manufacturing}.

%The commercial GlobalFoundries\textsuperscript{\texttrademark} (CF) quantum dot devices, as illustrated in Fig.~2(c), were fabricated using the 22FDX\textsuperscript{\textregistered} platform. Here, two FDSOI devices, having four high-k metal gate stacks, are parallel connected and integrated on the same integrated circuit (IC) die, having a common drain terminal. Each of the devices has a 6 nm silicon channel, 20 nm buried oxide and ESD-protected top gates. While the dots are formed under G1 and G2 gates, the gated spacer, together with the ungatedregions, controls the tunnel barriers. An n+ doped silicon layer below the buried oxide acts as a back gate, which also controls the interdot tunnel coupling. In the selective back-gate device (SB), the n+ doped silicon region is limited to the dotgates region, whereas in the full back-gate (FB) device, this layer extends throughout the entire device geometry. Further details of this device are available at {\cite{ajit2025}}

%All of the hallbar measurements were performed using standard four-terminal low-frequency lock-in techniques with an AC excitation voltage of 100~$\mu$V in a pumped Helium system with a base temperature of 1.5~K.In the quantum transport measurements, a particular carrier density was maintained by the top gate voltage. On the other hand, all of the quantum dot characterisation and charge noise spectroscopy measurements were performed in a liquid helium system at ~4K. A PicoScope 4824A was used to record the time domain current noise data. 

All Hall-bar measurements were performed using standard four-terminal, low-frequency (133 Hz) lock-in techniques with an AC excitation voltage of 100~$\mu$V in a pumped helium system with a base temperature of $\sim$1.4~K. For the quantum transport measurements, the carrier density was controlled via the applied top-gate voltage. All quantum dot characterization and charge-noise spectroscopy measurements were conducted in a liquid helium system at $\sim$4.2~K. Time-domain current noise data were recorded using a PicoScope~4824A oscilloscope.

\section{Results and Discussions}\label{sec2}

%(d) Extracted Dingle ratios, $\tau_{\mathrm{t}}/\tau_{\mathrm{q}}$, from representative devices corresponding to the four configurations shown in panel (c), providing insight into dominant scattering mechanisms and the quality of the quantum lifetime.

\subsection{Gate-Stack-Dependent Transport in Hall-Bar Devices}

%Figure 1 (b)$-$(c) summarizes peak mobility comparison of Hallbar devices fabricated with different high-$\kappa$ oxides and gate metals. In Fig. 1(b), Al$_2$O$_3$ film was deposited via ALD varying different oxidants and deposition temperature. It is reported that, using Al$_2$O$_3$ as an intermediate layer between SiO$_2$ and gate metal can help to reduce the dewetting and cryogenic stain effects~\cite{A_fabrication_guide_for_planar_silicon_quantum_dot_heterostructures, Palladium_gates_for_reproducible_quantum_dots_in_silicon}. Besides, it an be used as intergate dielectric~\cite{ Palladium_gates_for_reproducible_quantum_dots_in_silicon}.However, Nichol \textit{et al.} found that in Si/SiGe quantum dots, the noise increased with the Al$_2$O$_3$ gate oxide, consistent with more defects in the oxide producing more noise~\cite{Connors2019}. Thats why, initially we tried to optimize the Al$_2$O$_3$ deposition condition to improve the film quality. 

Figures~\ref{Figure 1_Chargenoise}(b)--(c) summarize the peak mobility comparison of Hall-bar devices fabricated with different high-$\kappa$ gate oxides and gate metals. In Fig.~\ref{Figure 1_Chargenoise}(b), Al$_2$O$_3$ films were deposited by ALD using different oxidants and deposition temperatures. Previous studies have shown that introducing Al$_2$O$_3$ as an intermediate layer between SiO$_2$ and the gate metal can mitigate metal dewetting and reduce cryogenic strain effects~\cite{A_fabrication_guide_for_planar_silicon_quantum_dot_heterostructures,Palladium_gates_for_reproducible_quantum_dots_in_silicon}. In addition, Al$_2$O$_3$ is also employed as an inter-gate dielectric in multilayer gate-stack architectures~\cite{Palladium_gates_for_reproducible_quantum_dots_in_silicon}. However, Connors \textit{et al.} reported that, in Si/SiGe quantum dots, the introduction of an Al$_2$O$_3$ gate oxide led to increased charge noise, consistent with a higher density of oxide-related defects~\cite{Connors2019}. Motivated by this observation, we first sought to optimize the ALD growth conditions of Al$_2$O$_3$ to improve film quality before incorporating it into nanoscale quantum dot devices.

The growth of thin films by ALD is based on sequential, self-limiting surface reactions between gaseous precursors and the substrate, with each reactant introduced alternately in a non-overlapping manner. For Al$_2$O$_3$ deposition, trimethylaluminum (TMA) and H$_2$O are the most commonly used metal precursor and oxidant, respectively~\cite{Puurunen2005,George2010}. Prior work has established that hydrogen incorporated during ALD Al$_2$O$_3$ growth can become mobile during post-deposition annealing and participate in the passivation of silicon dangling bonds at the Si/oxide interface. Isotope-tracing studies using deuterium have provided evidence for such hydrogen transport and interfacial reactions, offering a means to probe the underlying kinetics of defect passivation~\cite{Dingemans2010_APL,A_fabrication_guide_for_planar_silicon_quantum_dot_heterostructures}. On this basis, the use of D$_2$O as an ALD oxidant is investigated here as an exploratory approach to examine whether modified hydrogen incorporation and transport can influence interfacial properties relevant to SiMOS quantum devices. In addition to precursor choice, deposition temperature plays a crucial role in determining film quality. Each precursor exhibits a characteristic temperature window in which it remains reactive without condensation or thermal decomposition. For Al$_2$O$_3$ ALD, this window spans from 25~$^{\circ}$C~\cite{Ye2005} to $\sim$360~$^{\circ}$C~\cite{Fei2015}, with linear growth typically observed between 115~$^{\circ}$C and 300~$^{\circ}$C~\cite{Guder2012,Puurunen2005}. At higher temperatures, TMA undergoes thermal decomposition above $\sim$300~$^{\circ}$C, leading to deviations from ideal self-limiting ALD behavior~\cite{Puurunen2001}.

Figure~\ref{Figure 1_Chargenoise}(b) compares the peak mobility of Al$_2$O$_3$-based Hall-bar devices deposited using two different oxidants (H$_2$O and D$_2$O) at two deposition temperatures (200~$^{\circ}$C and 300~$^{\circ}$C). No significant dependence of the peak mobility on the choice of oxidant was observed; for both deposition temperatures, devices grown with H$_2$O and D$_2$O exhibited comparable mobility values. In contrast, a clear dependence on deposition temperature was evident. Devices fabricated at 300~$^{\circ}$C consistently showed higher peak mobilities for both oxidants compared to their 200~$^{\circ}$C counterparts. This enhancement in mobility is attributed to denser and more compact Al$_2$O$_3$ films formed at higher deposition temperatures, which are associated with reduced carbon impurities and fewer oxygen-related defects~\cite{Batra2015,Kim2022}. In addition, previous studies have reported a reduction in oxide trap density at higher ALD growth temperatures, further contributing to improved transport properties~\cite{Rahman2020}.

Figure~\ref{Figure 1_Chargenoise}(c) presents a comparison of the peak mobility for Hall-bar devices incorporating two high-$\kappa$ gate oxides (Al$_2$O$_3$ and HfO$_2$) and two gate metals (Al and TiPd), while the associated percolation threshold density, $n{_\mathrm{p}}$, is shown in Fig.~\ref{Figure 1_Chargenoise}(d). For this comparison, the optimized Al$_2$O$_3$ deposition conditions identified in Fig.~\ref{Figure 1_Chargenoise}(b) were used. A reduction in peak mobility was observed for the SiO$_2$/Al$_2$O$_3$ stack compared with the single SiO$_2$ case. This degradation is attributed to the formation of an interfacial dipole layer at the SiO$_2$/Al$_2$O$_3$ interface, which acts as a source of remote Coulomb scattering~\cite{Shimura_2016}.

HfO$_2$ is another extensively studied high-$\kappa$ dielectric that is commonly deposited by ALD~\cite{Jing2024}. It offers a moderate bandgap ($\sim$5.7~eV) with a higher dielectric constant ($\kappa \approx 20$--25) compared to Al$_2$O$_3$, which has a relatively lower dielectric constant ($\kappa \approx 6$--9)~\cite{Rahman2021}. Moreover, Si/SiO$_2$/HfO$_2$ gate stacks have been reported to exhibit promising carrier transport properties in MOSFET devices~\cite{Nunomura_2023}. Motivated by these advantages, we also evaluated the suitability of HfO$_2$ for quantum dot devices. Interestingly, the SiO$_2$/HfO$_2$ stack exhibited peak mobility values comparable to those of devices with a single SiO$_2$ gate oxide. This behaviour can be understood in terms of defect passivation mechanisms. Oxygen vacancies are widely recognised as the dominant intrinsic defects in HfO$_2$ films~\cite{Gavartin2006}. Nunomura \textit{et al.} reported that FGA enables hydrogen diffusion that passivates oxygen-vacancy and impurity-related defects in both HfO$_2$ and SiO$_2$ layers~\cite{Nunomura_2023}. Furthermore, owing to the smaller ionic radius of Al compared to Hf, Al atoms can diffuse into pinhole regions beneath the HfO$_2$ layer, forming a more continuous and compact Al-doped HfO$_2$ film~\cite{Zheng2015,Park2006}. Al incorporation has also been shown to stabilize the tetragonal phase of HfO$_2$, which possesses a higher dielectric constant~\cite{Lee2008}. In the present devices, the Al gate metal serves as a natural source of Al atoms, facilitating this beneficial doping effect.

Replacing the Al gate with a TiPd gate, another commonly used gate metal~\cite{Palladium_gates_for_reproducible_quantum_dots_in_silicon}, on the SiO$_2$/Al$_2$O$_3$ stack resulted in an approximately threefold reduction in electron mobility compared to devices employing Al as the gate metal. This degradation can be attributed, in part, to stronger strain-induced perturbations of the silicon conduction band introduced by Pd relative to Al, as previously reported~\cite{Alternatives_to_aluminum_gates_for_silicon_quantum_devices_Defects_and_strain}. Such strain modifies the local conduction-band edge at the atomic scale, distorting the electrostatic potential landscape and thereby enhancing carrier scattering and suppressing transport. Consistent with this interpretation, a substantially higher percolation threshold density was extracted for the TiPd-gated device, $n_{\mathrm{p}} = (7.013 \pm 0.053)\times10^{11}\,\mathrm{cm^{-2}}$, compared to $n_{\mathrm{p}} = (2.43 \pm 0.14)\times10^{11}\,\mathrm{cm^{-2}}$ for the SiO$_2$/Al$_2$O$_3$ stack with an Al gate as shown in Fig.~\ref{Figure 1_Chargenoise}(d). The percolation threshold density was obtained by fitting the carrier-density-dependent longitudinal conductivity, $\sigma_{\mathrm{XX}}$, to a semiclassical percolation-driven metal--insulator transition model, $\sigma_{\mathrm{XX}} \propto (n - n_{\mathrm{p}})^p$, with the critical exponent fixed at $p = 1.31$, as expected for a two-dimensional system~\cite{Observation_of_percolation_induced_two_dimensional_metal_insulator_transition_in_a_Si_MOSFET, Annealing_shallow_Si_SiO2_interface_traps_in_electron_beam_irradiated_high_mobility_metal_oxide_silicon_transistors}. Further details about the percolation density calculation process is described in Refs.~\cite{Quantum_Transport_Properties_of_Industrial, High_mobility_SiMOSFETs_fabricated_in_a_full_300_mm_CMOS_process}. The elevated $n_{\mathrm{p}}$ thus provides quantitative evidence for increased disorder in Pd-gated devices.

Beyond strain-related disorder, an additional mechanism that may further contribute to the reduced mobility in Pd-based devices is hydrogen uptake during the FGA process. During FGA, hydrogen typically diffuses through the oxide and passivates dangling bonds at the Si/SiO$_2$ interface, leading to improved carrier mobility. However, Pd exhibits a strong affinity for hydrogen, particularly at elevated temperatures~\cite{The_role_of_palladium_in_a_hydrogen_economy,Path_and_mechanism_of_hydrogen_absorption_at_Pd(100)}. Hydrogen readily permeates the Pd film, forming palladium hydride (PdH$_x$) and accumulating at the metal--oxide interface, where polarization effects induce an interfacial dipole that modifies the device electrostatics~\cite{Hydrogen_interaction_with_platinum_and_palladium_metal–insulator–semiconductor_device,Hydrogen_sensitive_mos-structures_Part_1_Principles_and_applications}. Moreover, hydrogen-induced interface states have been reported to enhance carrier scattering and degrade interface quality~\cite{Mechanism_of_the_formation_of_hydrogen‐induced_interface_states_for_Pt/silicon_oxide/Si_metal–oxide–semiconductor_tunneling_diodes,The_effect_of_hydrogen_and_carbon_monoxide_on_the_interface_state_density_in_MOS_gas_sensors_with_ultra-thin_palladium_gates}. Consistent with this picture, Pd-gated devices exhibit a significantly higher scattering charge density, $N_{\mathrm{C}} = (29.7 \pm 0.08)\times10^{10}\,\mathrm{cm^{-2}}$, compared to Al-gated devices, $N_{\mathrm{C}} = (4.83 \pm 0.17)\times10^{10}\,\mathrm{cm^{-2}}$, extracted from Kruithof--Klapwijk--Bakker fitting~\cite{Temperature_and_interface_roughness_dependence_of_the_electron_mobility_in_high_mobility_Si(100)_inversion_layers_below_4.2_K,Temperature_dependence_of_the_conductivity_for_the_two_dimensional_electron_gas_Analytical_results_for_low_temperatures}. In addition, Ti can act as an oxygen scavenger by extracting oxygen from the underlying SiO$_2$, thereby generating oxygen vacancies that further enhance scattering~\cite{High_mobility_SiMOSFETs_fabricated_in_a_full_300_mm_CMOS_process}. The observed mobility reduction can therefore be understood as the combined consequence of strain-induced disorder, hydrogen-related defect formation, and oxygen-vacancy-mediated scattering. Finally, magneto-transport measurements were performed on four representative devices corresponding to the configurations shown in Fig.~\ref{Figure 1_Chargenoise}(c), with additional details provided in Appendix~\ref{secA1}.

\subsection{Charge-Noise Analysis in the Many-Electron Regime}

\begin{figure}
\centering
\includegraphics[width=1\textwidth]{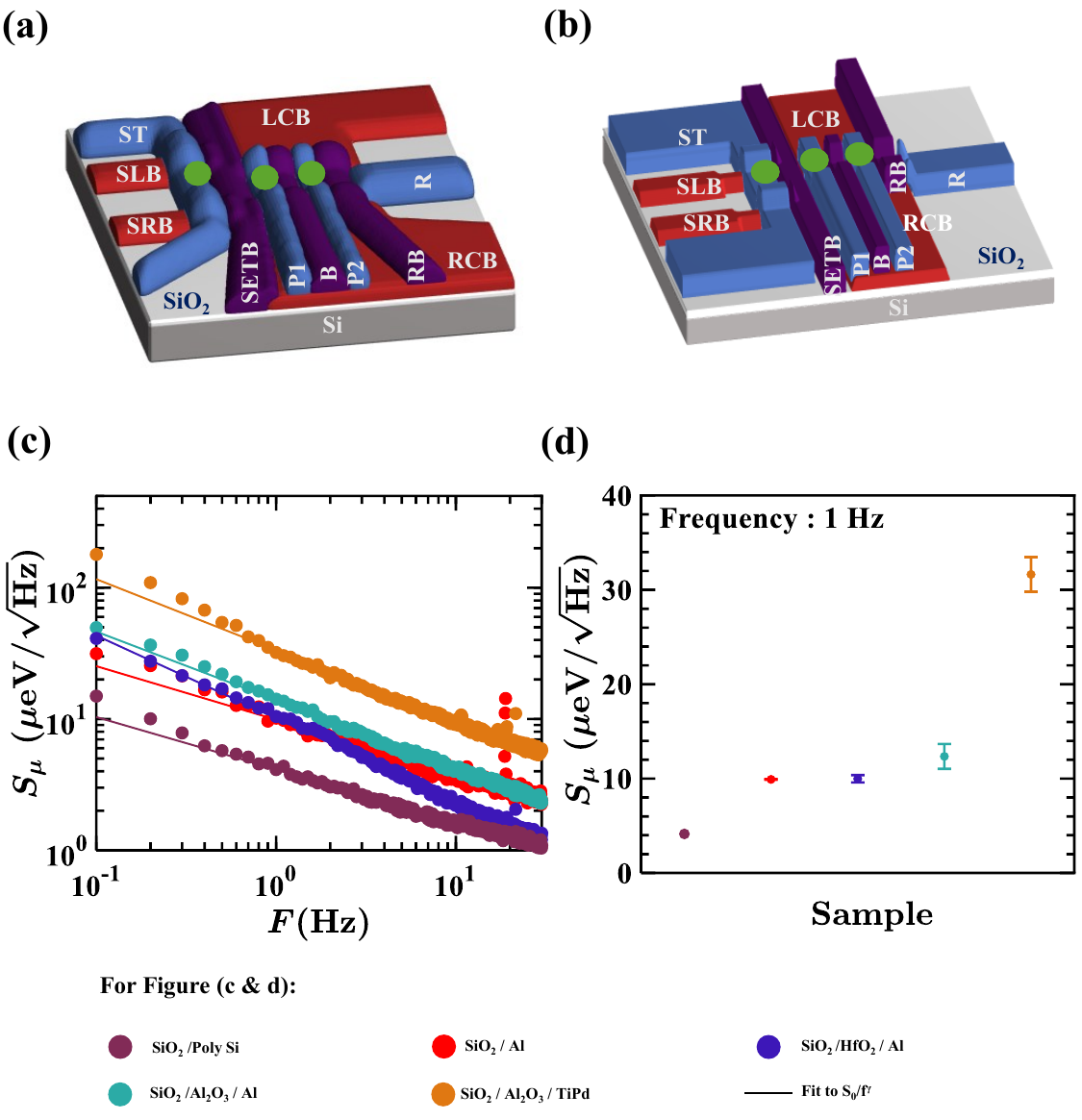}
\caption[\textbf{Quantum dot device architectures and charge-noise characterization}] {
\textbf{Quantum dot device architectures and charge-noise characterization} 
\textbf{(a)} Double quantum dot device fabricated in the UNSW cleanroom using overlapping \emph{metal gate electrodes} defined by a lift-off process.
\textbf{(b)} Double quantum dot device fabricated on a 300~mm wafer at imec using a CMOS-compatible process, employing \emph{etched poly-Si gate electrodes}.
In both schematics, the red gates represent the first-layer gates, the blue gates denote the second-layer gates, and the purple gates correspond to the third-layer gates. The thermally grown SiO$_2$ is shown in white, while the silicon substrate is illustrated in grey. Three quantum dots are formed beneath the ST (named as SET dot), P1, and P2 gates; however, for the analysis presented in panels (c) and (d), only the charge-noise data from the SET are considered.
\textbf{(c)} Charge-noise spectra measured from the SET dot of representative devices from the four device types shown in Fig.~\ref{Figure 1_Chargenoise}(c), with solid curves indicating power-law fits of the form $S_{\mathrm{0}}/f^{\gamma}$.
\textbf{(d)} Charge-noise amplitudes extracted at 1~Hz and 4~K for all measured devices, obtained from the power-law fits to the spectra shown in panel (c). For each device category, measurements were performed on at least two nominally identical devices, except for the poly-Si–gated device, where a single device was measured. The error bars represent the standard deviation of the measured values across devices within each configuration.
}
\label{Figure 2_Chargenoise}
\end{figure}

%Figure 2 shows the schematic of double dot devices use for the charge noise analysis. Figure 2(a) show the device fabricated  at UNSW clean room and Fig. 2(b) show the Diraq device fabricated at imec. In both devices, three overlapping layer of gates are formed depicted using same colours in both cases. Double dots are formed under the plunger gates P1 and P2 while B controls the interdot coupling. R is the reserviour gate while RB is the resrviour barrier. In both cases, a Single eletron transitor (SET) has formed under the ST gate , while SLB and SRB are the barrier gates, as a sensor dot. For the charge noise analysis presented in Fig. 2, we focused on the SET charge noise,measured in the many electron regime, as it has been shown to well represent the qubit noise`\cite{Struck2020, Connors2022} and typically the noise figure at 1Hz is used as the metric to benchmark between different oxides and material platforms.

Figure~\ref{Figure 2_Chargenoise} illustrates the double quantum dot device architectures used for the charge-noise measurements. Figure~\ref{Figure 2_Chargenoise}(a) shows a device fabricated in the UNSW clean room (UC), while Fig.~\ref{Figure 2_Chargenoise}(b) presents the device fabricated at imec. In both cases, three overlapping layers of electrostatic gates are defined and depicted using consistent colour coding for clarity. Double quantum dots are formed beneath the plunger gates P1 and P2, with gate B controlling the interdot tunnel coupling. R denotes the reservoir gate, while RB corresponds to the reservoir barrier that controls the coupling of the P2 dot to the electron reservoir. In addition, a single-electron transistor (SET) is formed beneath the ST gate, serving as a charge sensor, with SLB and SRB acting as the corresponding barrier gates. For the charge-noise analysis presented in Fig.~\ref{Figure 2_Chargenoise}, we focus exclusively on noise measured from the SET operated in the many-electron regime. Charge noise extracted from the SET under these conditions has been shown to provide a reliable proxy for the electrostatic noise environment relevant to spin-qubit operation~\cite{Struck2020, Connors2022}. Accordingly, the SET
charge-noise amplitude evaluated at 1~Hz is used as a quantitative metric to benchmark and compare noise performance across different gate oxides and material platforms.

In a quantum dot, charge noise appears as temporal fluctuations of the dot's electrochemical potential arising from coupling to the surrounding electrostatic environment. Rather than interacting with a single bistable defect, the dot is typically influenced by an ensemble of fluctuators that collectively contribute to the potential noise. Experimentally, these fluctuations are detected as current variations when the dot is biased on the slope of a Coulomb peak, where the conductance is most sensitive to changes in the local electric field. Operating the device at this point of maximum transconductance efficiently converts potential fluctuations into measurable current noise. Accordingly, when the SET is biased on the flanks of a Coulomb peak, the resulting current-noise spectrum is recorded and subsequently converted into equivalent quantum-dot potential fluctuations using the independently calibrated transconductance and gate lever arm. Further details of the charge-noise extraction procedure are provided in Appendix~\ref{secA2}.

Figure~\ref{Figure 2_Chargenoise}(c) presents the charge-noise spectra measured from the SET dots of the UC and RTF devices, together with power-law fits of the form $S_{\mathrm{0}}/f^{\gamma}$ (solid lines). For the UC platform, four representative devices, fabricated using different combinations of high-$\kappa$ gate oxides and gate metals, as summarized in Fig.~\ref{Figure 1_Chargenoise}(c), are included to assess the dependence of charge noise on gate-stack composition. For each device category, the spectrum shown corresponds to the representative device exhibiting the lowest charge noise, plotted alongside the RTF device for comparison.

Inspection of the spectra reveals that all devices exhibit a clear power-law dependence, $S(f)\propto 1/f^{\gamma}$, over the primary fitting range, where $\gamma$ is the power-law exponent. Within the standard two-level fluctuator (TLF) framework, such power-law behaviour is consistent with charge noise arising from a spatially distributed ensemble of fluctuators. If the fluctuators are distributed log-uniformly in their characteristic switching rates, a power-law exponent $\gamma \approx 1$ is expected. In contrast, the extracted exponents in this study lie in the range $\gamma = 0.4$--$0.7$, indicating power-law charge noise with sub-$1/f$ scaling. Within the Dutta--Horn formalism, such sub-$1/f$ behaviour arises naturally from an ensemble of thermally activated TLFs with a non-uniform distribution of activation energies and switching rates, without implying a change in the underlying noise mechanism~\cite{DuttaHorn1981, Connors2019}. Deviations from $\gamma=1$ have also been reported in charge-noise measurements of Si-based quantum devices~\cite{Dial2013, Connors2019, Connors2022, Chanrion2020}. In several devices, the measured noise at frequencies below $\sim$0.8~Hz lies above the extrapolated power-law fit. Although the spectra remain consistent with power-law charge noise over the main fitting range (1 Hz to 30 Hz), the lowest-frequency data deviate from a single power-law description. These deviations are not modelled separately, and the fit parameters are extracted exclusively from the frequency window where robust scaling behavior is observed; a detailed analysis of the lowest-frequency contributions is beyond the scope of this work.

Figure~\ref{Figure 2_Chargenoise}(d) summarizes the extracted charge-noise amplitudes at 1~Hz for all device configurations shown in Fig.~\ref{Figure 2_Chargenoise}(c), with error bars representing the standard deviation across multiple devices (at least two) for each configuration. A clear qualitative correlation is observed between charge noise and Hall-bar transport properties: devices with higher peak mobility exhibit lower charge noise. Among all devices measured, the RTF device employing a poly-Si gate on a single SiO$_2$ gate oxide exhibits the lowest charge-noise amplitude, $S_{\mathrm{\mu}}^{1/2}(1~\mathrm{Hz}) = 4.14~\mu\mathrm{eV}/\sqrt{\mathrm{Hz}}$. Compared to its UC counterpart employing an Al gate, this reduced noise level is consistent with several structural and materials differences. These include the thicker gate oxide (12~nm versus 8~nm), which increases the separation between the quantum dot and the oxide interfaces, as well as extensive gate-stack optimization and reduced strain associated with the use of poly-Si gates~\cite{Low_charge_noise_quantum_dots_with_industrial_CMOS_manufacturing}. In their recent study, Elsayed \textit{et al.} reported that increasing the SiO$_2$ thickness from 8~nm to 12~nm led to a substantial improvement in peak mobility, from $17.5\times10^{3}$ to $30\times10^{3}$~cm$^{2}$V$^{-1}$s$^{-1}$, accompanied by an increase in the Dingle ratio from $\sim$1 to $\sim$3~\cite{High_mobility_SiMOSFETs_fabricated_in_a_full_300_mm_CMOS_process, Low_charge_noise_quantum_dots_with_industrial_CMOS_manufacturing}. They further demonstrated that comprehensive gate-stack optimization resulted in charge-noise levels lower than previously reported values~\cite{Stuyck2021}. In addition, their earlier work showed that replacing poly-Si gates with TiN led to mobility degradation due to strain and oxygen-vacancy-related effects~\cite{High_mobility_SiMOSFETs_fabricated_in_a_full_300_mm_CMOS_process}. Taken together, these observations are consistent with the exceptionally low charge noise measured in the RTF device. Similar trends have also been observed between RTF and UC devices in Ref.~\cite{ajit2025}. 

For the UC devices employing Al gates, the dependence of charge noise on gate oxide composition closely follows the trends observed in transport. Devices with a SiO$_2$/HfO$_2$ gate stack exhibit charge-noise levels comparable to those of single SiO$_2$ devices, whereas devices incorporating a SiO$_2$/Al$_2$O$_3$ stack display noticeably higher noise. This behavior is consistent with the electrical transport results in Fig.~\ref{Figure 1_Chargenoise}(c), where reduced mobility was observed for the SiO$_2$/Al$_2$O$_3$ stack. In this case, dipole-like TLFs at the SiO$_2$/Al$_2$O$_3$ interface are expected to contribute both to enhanced charge noise and to remote Coulomb scattering, as previously reported for similar systems~\cite{Connors2019}. In contrast, for the SiO$_2$/HfO$_2$ stack, the comparable mobility and noise levels are consistent with improved film quality and defect passivation, potentially aided by Al incorporation during processing, as discussed earlier.

Finally, devices employing TiPd gates consistently exhibit the highest charge-noise levels among all configurations studied. This trend mirrors their substantially reduced mobility and elevated percolation threshold density. The increased charge noise is therefore consistent with enhanced disorder arising from a combination of hydrogen-related defect formation during FGA and oxygen-vacancy-mediated scattering associated with the Ti adhesion layer. Overall, the strong correspondence between mobility degradation and increased charge noise across all device platforms indicates that material- and interface-induced disorder plays a central role in determining the low-frequency electrostatic noise environment in these systems.

\begin{figure}
\centering
\includegraphics[width=0.8\textwidth]{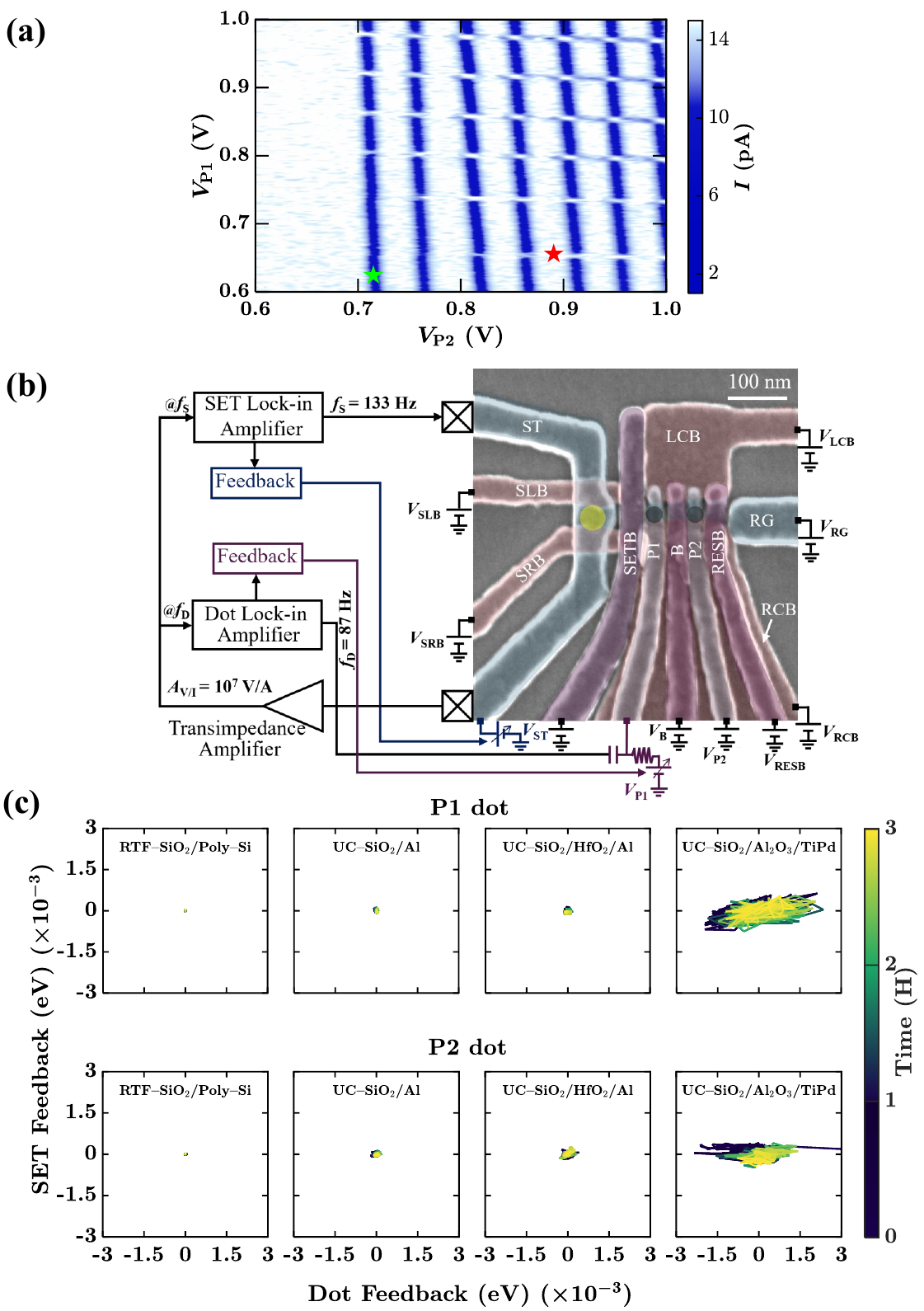}

\caption[\textbf{Charge-stability mapping and Dot--SET feedback characterization of quantum dot devices.}]{
\textbf{Charge-stability mapping and Dot--SET feedback characterization of quantum dot devices.}
\textbf{(a)} Charge-stability diagram of the RTF device, identified as the lowest charge-noise performer, plotted as a function of the plunger-gate voltages $V_{\mathrm{P1}}$ and $V_{\mathrm{P2}}$. The horizontal and vertical transition lines correspond to electron loading in the P1 and P2 quantum dots, respectively, down to the last occupied electron. The star symbols mark the transition points where the dot signals are held fixed for operation of the dual-feedback control system. 
\textbf{(b)} A dual-feedback control scheme is used to stabilize the operating currents of both the SET sensor and the quantum dot by dynamically adjusting their respective gate voltages. The recorded feedback corrections, $\Delta V_{\mathrm{ST}}^{(\mathrm{fb})}$ and $\Delta V_{\mathrm{P}}^{(\mathrm{fb})}$, quantify the electrostatic compensation required to maintain fixed operating points and are converted to energy units using independently calibrated gate lever arms.
\textbf{(c)} Comparison of dot--SET feedback stability maps from four representative devices incorporating different high-$\kappa$ oxides and gate-metal stacks, shown for the first-electron transition of the P1 and P2 dots. These were recorded continuously over a three-hour interval. The temporal drift of the operating point is visualized using the associated color scale. A larger spread or distortion of the stability map indicates increased charge instability and a noisier electrostatic environment.
}
\label{Figure 3_Chargenoise}
\end{figure}
\subsection{Dot–Sensor Charge Stability in Few-Electron Regime}

Charge-noise measurements based on the SET current are inherently performed in the many-electron regime of the sensor dot and therefore primarily characterize the electrostatic noise environment under these conditions. In contrast, spin-qubit operation requires quantum dots to be tuned to very low electron occupancy, typically from a single to only a few electrons, where the influence of electrostatic fluctuations may differ substantially. Consequently, charge-noise measurements performed at fixed sensor or dot currents do not fully capture the noise processes most relevant to qubit operation. To more closely emulate qubit-operating conditions, we instead probe charge-noise-induced fluctuations in a coupled dot–sensor system with the quantum dot tuned to the few-electron regime. This approach enables a more direct assessment of electrostatic potential fluctuations that limit qubit stability. Specifically, we employ the dynamically controlled charge-sensing technique introduced by Yang \textit{et al.}~\cite{Yang2011}, implementing a dual-feedback algorithm that stabilizes both the SET sensor current and the quantum-dot sensing signal at fixed operating points. The feedback voltages applied to the SET and dot plunger gates thus provide a real-time, qualitative measure of the electrostatic potential fluctuations experienced by each subsystem.

Figure~\ref{Figure 3_Chargenoise}(a) presents a charge-stability diagram of the P1 and P2 quantum dots in the RTF device, which exhibits the lowest charge noise and is therefore used as a reference. The diagram is obtained by sweeping the plunger-gate voltages $V_{\mathrm{P1}}$ and $V_{\mathrm{P2}}$ while keeping all other gates fixed. Distinct charge transitions associated with electron loading and unloading are clearly resolved down to the last electron. For the dot–SET stability analysis, we focus on the first-electron transition of both P1 and P2.

A dual-feedback measurement protocol was then implemented to stabilize both the SET sensor and the quantum dot at fixed operating points. The SET was tuned to the flank of a Coulomb peak, and a target sensor current $I_{\mathrm{S0}}$ was chosen at a point of high transconductance. Likewise, the quantum dot was biased on the flank of a charge transition signal, and a target dot sensing signal $I_{\mathrm{D0}}$ was selected. The sensor current, $I_{\mathrm{S}}(t)$, and dot sensing signal, $I_{\mathrm{D}}(t)$, were continuously monitored for several hours while feedback controllers dynamically adjusted the corresponding gate voltages to compensate for slow electrostatic drifts and charge rearrangements. Deviations from the operating points were defined as $i_{\mathrm{S}} = I_{\mathrm{S}} - I_{\mathrm{S0}}$ and $i_{\mathrm{D}} = I_{\mathrm{D}} - I_{\mathrm{D0}}$.

The SET current was stabilized using a second-order feedback controller, similar to previously reported dynamically controlled charge-sensing schemes, by applying an additional feedback voltage to the SET plunger gate $V_{\mathrm{ST}}$. In parallel, an independent feedback loop applied a compensating voltage to a designated dot gate $V_{\mathrm{P}}$ to maintain $I_{\mathrm{D}} = I_{\mathrm{D0}}$. For each measurement point, the steady-state feedback voltages required to maintain stability, $\Delta V_{\mathrm{ST}}^{(\mathrm{fb})}$ and $\Delta V_{\mathrm{P}}^{(\mathrm{fb})}$, were recorded. These voltages quantify the electrostatic compensation required to counteract slow charge rearrangements and environmental fluctuations. Using independently calibrated lever arms $\alpha_{\mathrm{S}}$ and $\alpha_{\mathrm{D}}$, the voltages were converted to equivalent energy shifts $\Delta E_{\mathrm{S}}$ and $\Delta E_{\mathrm{D}}$. Plotting $\Delta E_{\mathrm{D}}$ versus $\Delta E_{\mathrm{S}}$ yields a correlated energy–energy stability map, providing a dual-feedback analogue to conventional charge-sensing maps. A conceptual schematic of the dual-feedback scheme is shown in Fig.~\ref{Figure 3_Chargenoise}(b).

Figure~\ref{Figure 3_Chargenoise}(c) displays the dual-feedback dot–SET stability maps for the P1 and P2 dots, measured separately across multiple material systems incorporating different gate oxides and gate metals. Each map reflects a three-hour measurement during which both currents were actively stabilized. Each point indicates the simultaneous feedback compensation required to maintain the operating points. The spatial extent of the stability region, therefore, reflects the magnitude of the electrostatic potential fluctuations experienced by the coupled dot–sensor system: a smaller span signifies reduced charge noise, whereas a larger span indicates a noisier electrostatic environment.

As a reference, the RTF device—featuring a poly-Si CMOS gate stack—shows the smallest stability-map span for both dots, indicating superior charge stability. In contrast, devices with SiO$_2$/Al$_2$O$_3$/TiPd gate stacks display the largest spans, consistent with the strongest charge instability. These trends align quantitatively with the SET-based charge-noise amplitudes extracted in Fig.~\ref{Figure 2_Chargenoise}(d), demonstrating consistency between the two independent measurements. Although SiO$_2$/HfO$_2$/Al devices exhibit charge-noise amplitudes similar to SiO$_2$/Al devices, their dual-feedback stability maps exhibit a slightly larger spread, suggesting subtle differences in the spatial coupling or temporal dynamics of the dominant fluctuators. In several devices, the P2 dot shows a marginally larger span than the P1 dot, indicating dot-to-dot variability within the same device. Occasional spike-like features observed along the dot-feedback axis may be attributable to abrupt charge rearrangements associated with individual traps.

Overall, these measurements show that while the underlying charge-noise mechanism is common across devices, variations in the spatial distribution and capacitive coupling of fluctuators lead to distinct dot–sensor coupling geometries. The dual-feedback stability-map technique directly reveals these differences, providing a real-time, correlated view of electrostatic potential fluctuations in a coupled dot–sensor system.

\section{Conclusion}\label{sec2}

In summary, we demonstrated that charge noise in SiMOS quantum devices is strongly governed by gate-stack engineering. Transport measurements show that oxide stoichiometry, controlled through ALD growth temperature, plays a decisive role in enhancing mobility, while gate-metal choice critically impacts both mobility and noise. HfO$_2$ gate stacks benefit from defect passivation effects, whereas TiPd gates introduce additional disorder that degrades transport and increases charge noise. Consistent correlations between mobility and noise are observed across Hall-bar and quantum-dot measurements. Dual-feedback stability mapping shows that optimized gate stacks are associated with improved charge stability, providing useful insight into material choices for the design of low-noise silicon qubit devices.

\bmhead{Data Availability}
The data that support the findings of this study are available based on a reasonable request from the corresponding author.

%\bmhead{Supplementary information}

\bmhead{Acknowledgments}

We acknowledge support from the Australian Research Council (FL190100167, CE170100012, and IM230100396), the US Army Research Office (W911NF-17-1-0198, W911NF-23-10092), and the NSW Node of the Australian National Fabrication Facility(ANFF) at the University of New South Wales (UNSW). The views and conclusions contained in this document are those of the authors and should not be interpreted as representing the official policies, either expressed or implied, of the Army Research Office or the US Government. The US Government is authorised to reproduce and distribute reprints for Government purposes, notwithstanding any copyright notation herein. The authors acknowledge the facilities as well as the scientific and technical assistance of the Research and Prototype Foundry Core Research Facility at the University of Sydney, part of the NSW node of the NCRIS-enabled Australian National Fabrication Facility.  M.M.R. acknowledges the scholarship support from the Sydney Quantum Academy, Australia.

We thank Paul Steinacker, Daniel Schwienbacher, and Santiago Serrano for their technical help with measurement setup and cryogenic setup maintenance. We also thank Professor Alexander Hamilton, Jonathan Wendoloski, and Usama Ahsan for their suggestions on the mobility fit model and Dingle plot fitting. We also thank Henry Yang for his support in charge-noise measurement.

%\bmhead{Research Funding}

%This work is funded by the Australian Research Council (Grants Nos. FL190100167, CE170100012, and IM230100396) and the US Army Research Office ( Grant Nos. W911NF-17-1-0198, W911NF-23-10092). 

%All fabrication processes were done at the academic cleanrooms of the Australian National Fabrication Facility (ANFF) at the University of New South Wales (UNSW).

\bmhead{Author contributions}

M.M.R. carried out the Hall-bar fabrication and the nano-fabrication of the quantum dot devices under the guidance of K.W.C. and W.H.L. W.H.L., K.W.C., F.H., and N.D.S. designed the devices, and F.H. performed the micro-fabrication of the devices under the supervision of A.S.D. M.M.R., E.V., A.D., and V.C. performed the experiments and data analysis under the supervision of W.H.L., K.W.C., and T.T., with input from A.L., J.H.C., A.S., and A.S.D. A.L. and A.M. contributed to the experimental hardware and cryogenic setup, while J.Y.H., J.D.C., and J.H.C. contributed to the experimental software setup. J.H.C., S.Y., Y.K.L., and A.S. contributed to the Shubnikov–de Haas (SdH) measurements and analysis. M.M.R., K.W.C., E.V., T.T., A.D., C.C.E., J.H.C., A.L., A.S., A.S.D., and W.H.L. contributed to the discussion, interpretation, and presentation of the results. M.M.R. wrote the manuscript with contributions from all authors.

\bmhead{Competing Interests}
 A.S.D. is the CEO and a director of Diraq Pty Ltd. W.H.L., K.W.C, F.H., A.D., T.T., E.V., N.D.S., J.D.C, C.C.E., Y.K.L., J.H.C., A.L., A.S., and A.S.D. declare equity interest in Diraq Pty Ltd.

\begin{appendices}

\section{Quantum-transport and scattering mechanisms of representative SiMOS Hall bars}\label{secA1}

\begin{figure}[!htbp]
\centering
\includegraphics[width=1\textwidth]{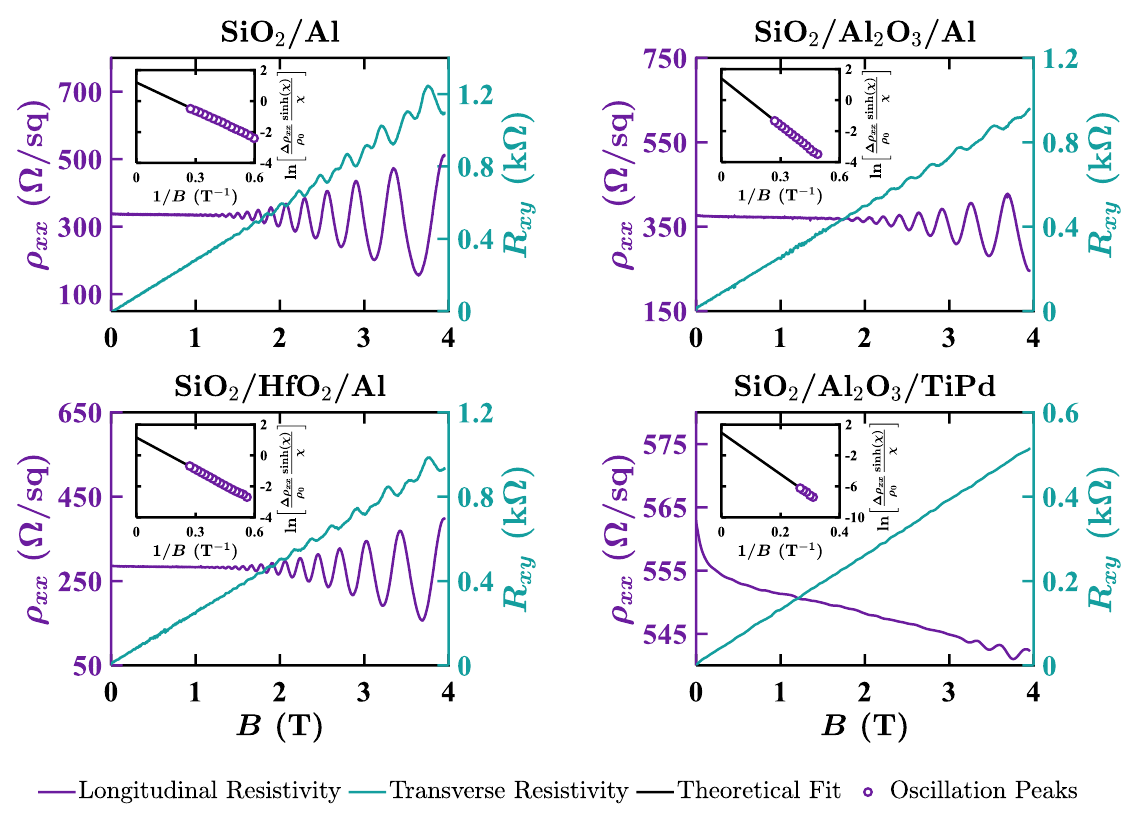}

\caption[\textbf{Magneto-transport characteristics of Hall-bar devices.}]{
\textbf{Magneto-transport characteristics of Hall-bar devices.}
Longitudinal resistivity $\rho_{\mathrm{XX}}$ and Hall resistance $R_{\mathrm{XY}}$ measured as functions of magnetic field B at 1.4~K for four Hall-bar devices incorporating different high-$\kappa$ oxides and gate-metal stacks.
\textit{Inset:} Dingle analysis of the Shubnikov--de Haas oscillations used to extract the quantum lifetime $\tau_{\mathrm{q}}$, with the solid line indicating the fit to the oscillation amplitudes.
}

\label{Figure 4_Chargenoise}
\end{figure}

To investigate the quantum transport behaviour associated with the two high-$\kappa$ gate oxides and two gate metals shown in Fig.~\ref{Figure 1_Chargenoise}(c), magneto-transport measurements were performed as a function of perpendicular magnetic field $B$ at a fixed carrier density, controlled by the top-gate voltage, for four representative devices. Figure~\ref{Figure 4_Chargenoise} presents the longitudinal resistivity, $\rho_{\mathrm{XX}}$ (purple), and the transverse (Hall) resistance, $R_{\mathrm{XY}}$ (cyan), measured as a function of magnetic field up to 3.9~T at 1.4~K for the four device configurations. Shubnikov--de Haas (SdH) oscillations became discernible above $\sim$1~T for the SiO$_2$/Al and SiO$_2$/HfO$_2$/Al devices, above $\sim$1.5~T for the SiO$_2$/Al$_2$O$_3$/Al device, and only above $\sim$3.2~T for the SiO$_2$/Al$_2$O$_3$/TiPd device, where the oscillation amplitude remained weak. In devices employing Al gates—most prominently the SiO$_2$/Al and SiO$_2$/HfO$_2$/Al stacks—the Hall resistance $R_{\mathrm{XY}}$ exhibited anomalous behavior, deviating from well-defined quantum Hall plateaus. Rather than evolving monotonically between successive quantized values, $R_{\mathrm{XY}}$ displayed a pronounced overshoot at lower magnetic fields before converging to the expected plateau, coincident with minima in $\rho_{\mathrm{XX}}$. This behavior is consistent with the quantum Hall resistance overshoot effect, attributed to transport through coexisting evanescent incompressible edge channels with different filling factors, and has previously been observed in Si-based two-dimensional electron systems such as Si-MOSFETs and Si/SiGe heterostructures~\cite{Quantum_Hall_resistance_overshoot_in_two-dimensional_(2D)_electron_gases,Transverse_“resistance_overshoot”_in_a_Si/SiGe_two-dimensional_electron_gas_in_the_quantum_Hall_effect_regime,Disorder-induced_features_of_the_transverse_resistance}. In contrast, no discernible quantum Hall plateaus were observed in the transverse resistance $R_{\mathrm{XY}}$ for devices with the SiO$_2$/Al$_2$O$_3$/TiPd gate stack, consistent with their significantly reduced mobility and enhanced disorder.

To identify the dominant scattering mechanisms, the Dingle ratio, $\tau_{\mathrm{t}}/\tau_{\mathrm{q}}$, was evaluated for each device. Using an effective mass of $m^{*}=0.19m_{e}$ and the mobility corresponding to the carrier density of the Hall measurements, the transport lifetime was calculated as $\tau_{\mathrm{t}}=\mu m^{*}/e$, where $e$ is the elementary charge. The extracted transport lifetimes were 0.952~ps, 0.701~ps, 0.921~ps, and 0.244~ps for the SiO$_2$/Al, SiO$_2$/Al$_2$O$_3$/Al, SiO$_2$/HfO$_2$/Al, and SiO$_2$/Al$_2$O$_3$/TiPd devices, respectively. Using the same effective mass, the quantum lifetime $\tau_{\mathrm{q}}$ was obtained from the slope of the Dingle plot shown in the inset of each panel by fitting the envelope of the SdH oscillation amplitudes, following the procedure described in Ref.~\cite{Low_charge_noise_quantum_dots_with_industrial_CMOS_manufacturing}. The extracted quantum lifetimes were 0.566~ps, 0.337~ps, 0.495~ps, and 0.126~ps for the same sequence of devices, yielding Dingle ratios of 1.68, 2.08, 1.86, and 1.94, respectively. The observed range of Dingle ratios indicates that carrier transport is predominantly limited by short-range, large-angle scattering, consistent with scattering centers located within or in close proximity to the two-dimensional electron gas, namely inside or near the channel region~\cite{Quantum_Transport_Properties_of_Industrial,High_mobility_SiMOSFETs_fabricated_in_a_full_300_mm_CMOS_process,Magnetotransport_studies_of_mobility_limiting_mechanisms_in_undoped_Si/SiGe_heterostructures}. The modest increase in the Dingle ratio for devices incorporating bilayer gate stacks further suggests the presence of an additional population of scattering centers, likely associated with the interface between the two oxide layers.

\section{Extraction process of charge noise }\label{secA2}

\begin{figure}[!htbp]
\centering
\includegraphics[width=0.8\textwidth]{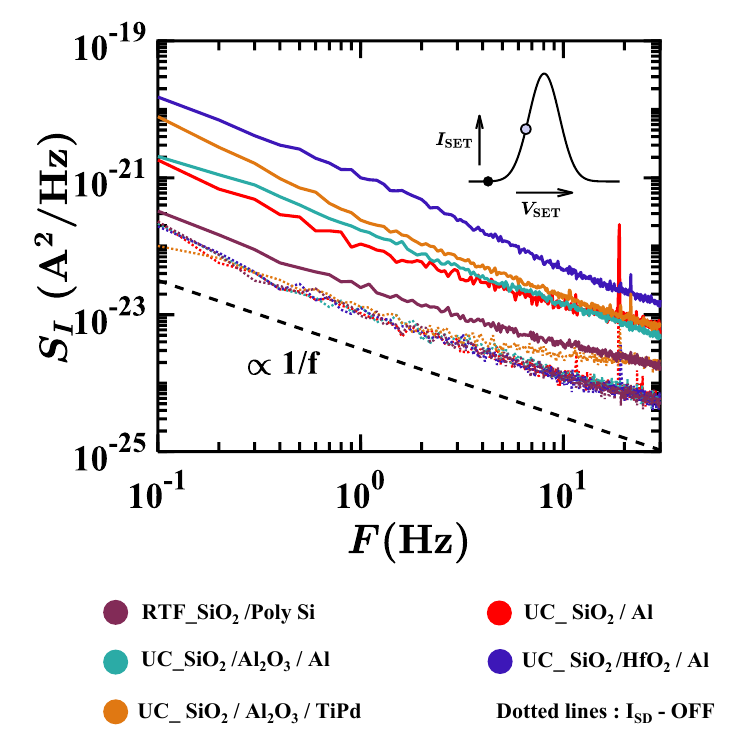}

\caption[\textbf{Current-noise spectroscopy of quantum dot devices.}]{
\textbf{Current-noise spectroscopy of quantum dot devices.}
Current-noise power spectral densities measured from five representative devices with different oxide stacks and gate metals, fabricated on the UC and RTF platforms. Solid curves show the noise spectra measured at the maximum of the $dI/dV$ transconductance, while dotted curves indicate the baseline noise measured with the devices in the OFF state (no source--drain current), as illustrated in the inset.
}

\label{Figure 5_Chargenoise}
\end{figure}

When a quantum dot is biased on a finite-slope region of a Coulomb blockade resonance, small perturbations of its electrochemical potential are transduced into measurable variations of the transport current. In this regime, fluctuations in the dot energy shift the position of the Coulomb peak relative to the fixed gate-voltage bias, thereby modulating the current. The dot electrochemical potential $\varepsilon$ is controlled by the gate voltage through the lever arm $\alpha$, such that a small energy fluctuation $\delta \varepsilon$ corresponds to an effective gate-voltage shift $\delta V_{\mathrm{G}} = \delta \varepsilon / \alpha$. Expanding the current to first order about the operating point yields
\begin{equation}
\delta I = \frac{dI}{dV_\delta V_{\mathrm{G}}}\,\delta V_\delta V_{\mathrm{G}} 
= \left(\frac{1}{\alpha}\frac{dI}{dV_\delta V_{\mathrm{G}}}\right)\delta \varepsilon ,
\end{equation}
where $V_G$ denotes the top-gate voltage applied to the quantum dot, which controls the dot electrochemical potential via the lever arm $\alpha$ and $dI/dV_{\mathrm{G}}$ is the local transconductance evaluated at the bias point. This linear relation, valid for fluctuations small compared to the Coulomb peak width, forms the basis for converting measured current noise into an equivalent charge-noise spectrum.

Charge-noise measurements were performed by operating the device on the flank of a selected Coulomb blockade peak, where the transconductance is maximal. The gate-voltage window spanning the peak was discretized into approximately 30 voltage setpoints covering the left flank, peak maximum, and right flank. At each setpoint, the drain current $I_{\mathrm{SD}}(t)$ was recorded over a duration of $T=10$~s and repeated $N=10$ times; the resulting time traces were averaged to suppress random acquisition noise. This full gate-voltage scan was repeated at least four times to ensure reproducibility.

For each averaged time trace, the current-noise power spectral density (PSD) was obtained via fast Fourier transform (FFT),
\begin{equation}
S_I(f)=\frac{2}{T}\,\bigl|\mathcal{F}\{ I_{\mathrm{SD}}(t)-\langle I_{\mathrm{SD}}\rangle \}\bigr|^2 ,
\end{equation}
where $\langle I_{\mathrm{SD}}\rangle$ denotes the time-averaged current and $\mathcal{F}\{\cdot\}$ represents the Fourier transform. The local sensitivity to electrostatic fluctuations was quantified by the magnitude of the transconductance,
\begin{equation}
g_{\mathrm{m}}(V_{\mathrm{G}})=\left|\frac{dI_{\mathrm{SD}}}{dV_{\mathrm{G}}}\right|.
\end{equation}
Only spectra acquired at high-sensitivity operating points near the maximum of $g_{\mathrm{m}}$ were retained for charge-noise extraction, while spectra measured at low-sensitivity points—including the peak maximum and far tails—were discarded. This selection ensures that the extracted noise is dominated by electrostatic potential fluctuations rather than reduced transconductance.

For each retained operating point, the current noise was converted to an equivalent gate-voltage noise using the linear-response relation
\[
\delta I = \left(\frac{dI_{\mathrm{SD}}}{dV_{\mathrm{G}}}\right)\delta V_{\mathrm{G}},
\]
yielding
\begin{equation}
S_{\mathrm{V}}(f)=\frac{S_{\mathrm{I}}(f)}{\left(\dfrac{dI_{\mathrm{SD}}}{dV_{\mathrm{G}}}\right)^2}.
\end{equation}
The selected spectra were truncated to a frequency range of 0.1-30 Hz and arithmetically averaged to obtain ensemble-averaged power spectral densities $\langle S_{\mathrm{I}}(f)\rangle$ and $\langle S_{\mathrm{V}}(f)\rangle$. The voltage noise was converted to charge noise in energy units using the independently calibrated lever arm, $\alpha$, obtained from Coulomb diamond measurements (ranging from 0.05 to 0.24 across devices),
\begin{equation}
S_{\mu}(f) = \alpha^2 \langle S_{\mathrm{V}}(f)\rangle,
\qquad
\sqrt{S_{\mu}(f)} = \alpha\,\sqrt{\langle S_{\mathrm{V}}(f)\rangle},
\end{equation}
where $\sqrt{S_{\mu}(f)}$ is reported as the charge-noise amplitude spectral density in units of $\mu\mathrm{eV}/\sqrt{\mathrm{Hz}}$. The charge-noise amplitude spectral density, $\sqrt{S_{\mu}(f)}$, was fitted to a power-law model,
\begin{equation}
\sqrt{S_{\mu}(f)}=\frac{S_0}{f^{\gamma}},
\end{equation}
from which the power-law exponent $\gamma$ and the charge-noise amplitude at 1~Hz were extracted.

To verify that the selected operating points corresponded to genuine device transport rather than the measurement floor, we additionally compared the mean (DC) current values at the retained high-sensitivity indices with the current measured when the device was in the OFF state (no source--drain current). Consistent with this validation, Fig.~\ref{Figure 5_Chargenoise} shows that the solid curves (measured at the maximum of $|dI/dV|$) lie above the dotted OFF-state traces, which quantify the baseline noise of the measurement setup.

\section{Dot-to-Dot charge noise variation in a device}\label{secA3}

\begin{figure}[!htbt]
\centering
\includegraphics[width=0.8\textwidth]{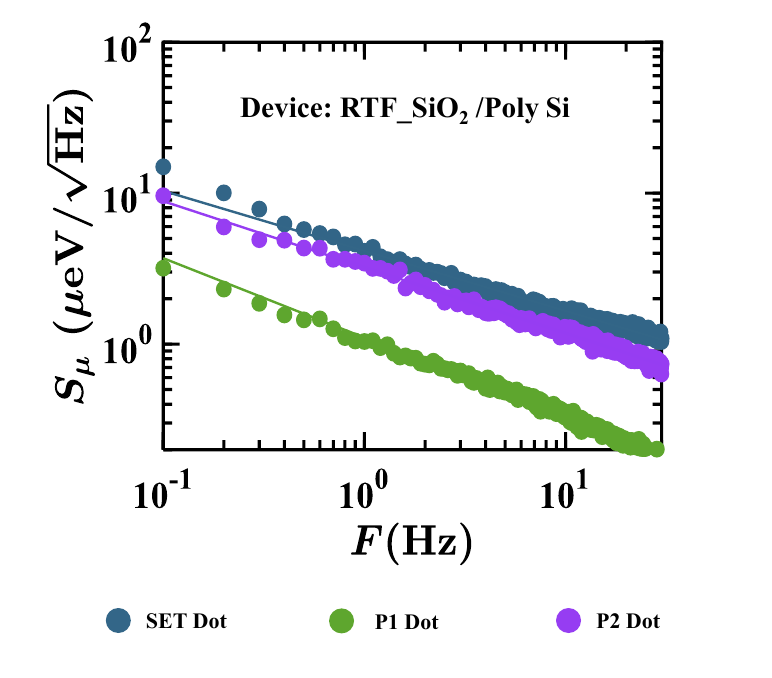}
\caption[\textbf{Charge noise spectra of quantum dots within the same device.}]{
\textbf{Charge noise spectra of quantum dots within the same device.}
Charge-noise spectra of quantum dots formed beneath the SET, P1, and P2 gates of the poly-Si--gated RTF device. For the SET dot, the transport current was measured between the source and drain contacts, whereas for the P1 and P2 dots, the current was measured between the drain contact and the reservoir ohmic.
}
\label{Figure 6_Chargenoise}
\end{figure}

In addition to comparing SET-based charge noise across different device architectures, we also investigated dot-to-dot variations in charge noise within a single device. Figure~\ref{Figure 6_Chargenoise} presents the charge-noise spectra measured from the P1, P2, and SET dots of the RTF device that exhibited the lowest overall noise in Fig.~\ref{Figure 2_Chargenoise}(d). A pronounced dot-to-dot variation in charge noise is observed. While the P2 dot exhibits a noise amplitude of $3.72~\mu\mathrm{eV}/\sqrt{\mathrm{Hz}}$, comparable to that of the SET dot ($4.14~\mu\mathrm{eV}/\sqrt{\mathrm{Hz}}$), the P1 dot shows a substantially lower value of $1.10~\mu\mathrm{eV}/\sqrt{\mathrm{Hz}}$, the lowest among all dots measured in this device. Such variations are consistent with the picture that individual quantum dots couple to distinct ensembles of two-level fluctuators, each characterized by a non-uniform distribution of activation energies and switching dynamics. Differences in the spatial proximity, density, and dynamics of these fluctuators can therefore lead to significant dot-to-dot differences in charge-noise amplitudes, even within the same device. Similar dot-dependent charge-noise variations have previously been reported in silicon quantum dot systems~\cite{Connors2019}.

\section{Dot--SET feedback stability map for second and third transitions}\label{secA4}

\begin{figure}[!htbp]
\centering
\includegraphics[width=0.9\textwidth]{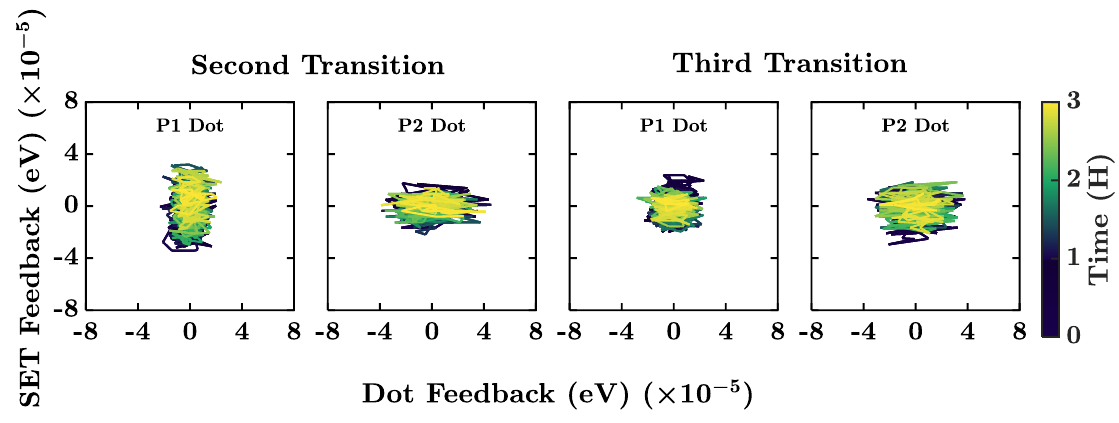}

\caption[\textbf{Dot--SET feedback stability maps of the poly-Si--gated RTF device.}]{
\textbf{Dot--SET feedback stability maps of the poly-Si--gated RTF device.}
Stability maps were recorded continuously over a three-hour interval at the second and third electron transitions of the P1 and P2 dots. The prolate structure observed in the stability map of the second-electron transition of the P1 dot indicates enhanced charge stability of the P1 dot relative to the SET charge sensor.
}
\label{Figure 7_Chargenoise}
\end{figure}

We additionally measured dot–sensor stability maps at the second and third electron transitions of the P1 and P2 dots in the RTF device to examine the dependence of charge stability on electron occupancy. No systematic correlation between electron number and the overall extent of the stability map was observed. For the P1 dot, a hollow stability region similar to that seen at the first-electron transition persists at the second transition, while the map becomes more compact at the third transition. In contrast, the P2 dot exhibits qualitatively similar stability-map shapes for both the second and third electron transitions, with no pronounced change in extent or anisotropy. These observations indicate that, within the few-electron regime investigated here, the dot–sensor stability is not strongly governed by electron number. Instead, the stability-map morphology appears to be dominated by the local electrostatic environment and the spatial distribution of nearby charge fluctuators, rather than by changes in the dot occupancy itself.

\end{appendices}

%\bibliography{sn-bibliography}% common bib file
%% if required, the content of .bbl file can be included here once the bbl is generated

\end{document}